\documentclass{article}

\usepackage{arxiv}

\usepackage[utf8]{inputenc} 
\usepackage[T1]{fontenc}    
\usepackage{hyperref}       
\usepackage{url}            
\usepackage{booktabs}       
\usepackage{amsfonts}       
\usepackage{nicefrac}       
\usepackage{microtype}      
\usepackage{lipsum}
\usepackage{natbib}

\title{Good proctor or ``Big Brother''? AI Ethics and Online Exam Supervision Technologies}

\author{Simon Coghlan$^{1,3}$, Tim Miller$^{1,3}$, and Jeannie Paterson$^{2,3}$\\
$^1$School of Computing and Information Systems\\
$^2$Melbourne Law School\\
$^3$The Centre for AI and Digital Ethics\\
The University of Melbourne, Australia\\
 \{simon.coghlan,tmiller,jeanniep\}@unimelb.edu.au
}

\begin{document}
\maketitle

\begin{abstract}
This article philosophically analyzes online exam supervision technologies, which have been thrust into the public spotlight due to campus lockdowns during the COVID-19 pandemic and the growing demand for online courses. Online exam proctoring technologies
purport to provide effective oversight of students sitting online exams,
using artificial intelligence (AI) systems and human invigilators to
supplement and review those systems. Such technologies have alarmed some
students who see them as `Big Brother-like', yet some universities
defend their judicious use. Critical ethical appraisal of online
proctoring technologies is overdue. This article philosophically analyzes
these technologies, focusing on the ethical concepts of academic
integrity, fairness, non-maleficence, transparency, privacy, respect for
autonomy, liberty, and trust. Most of these concepts are prominent in
the new field of AI ethics and all are relevant to the education
context. The essay provides ethical considerations that educational
institutions will need to carefully review before electing to deploy and
govern specific online proctoring technologies.
\end{abstract}


\section{Introduction}

Recently, online exam supervision technologies have been thrust into the public spotlight due to the growing demand for online courses \citep{ginder_enrollment_2019} and lockdowns during the COVID-19 pandemic \citep{flaherty_online_2020}. While educational institutions can supervise remote exam-takers simply by watching live online video (e.g. via Zoom), an evolving range of online proctoring (OP) software programs offer more sophisticated, scalable, and extensive monitoring functions, including both human-led and automated remote exam supervision. Such technologies have generated confusion and controversy, including vigorous student protests \citep{white_creepy_2020}. Some universities have dug in against criticism, while others have outright rejected the technologies or have retreated from their initial intentions to use them \citep{white_creepy_2020}. At the root of disagreement and debate between concerned students and universities are questions about the ethics of OP technologies. This essay explores these ethical questions. By doing so, it should assist students and educators in making informed judgements about the appropriateness of OP systems, as well as shining a light on an increasingly popular digital technology application.

OP software platforms, which first emerged in 2008 \citep{proctoru_proctoru_2020}, are now booming. A 2020 poll found that 54\% of educational institutions now use them \citep{grajek_educause_2020}. Increasingly, OP software contains artificial intelligence (AI) and machine learning (ML) components that analyse exam recordings to identify suspicious examinee behaviours or suspicious items in their immediate environment. OP companies, which can make good profits from their products \citep{chin_exam_2020}, claim that automating proctoring increases the scalability, efficiency, and accuracy of exam supervision and the detection of cheating. These features have an obvious attraction for universities, some of which believe the benefits of OP technologies outweigh any drawbacks. However, the complexity and opacity of OP technologies, especially their automated AI functions \citep{hagendorff_ethics_2020}, can be confusing. Furthermore, some (though not all) students complain of a ``creepy'' Big Brother sense of being invaded and surveiled \cite{hubler_keeping_2020}. Predictably, some bloggers are instructing students how to bluff proctoring platforms \citep{binstein_how_2015}.

Scholars have just begun exploring remote and automated proctoring from a range of perspectives, including pedagogical, behavioral, psychological, and technical perspectives \citep{asep_design_2019,cramp_lessons_2019,gonzalez-gonzalez_implementation_2020}. Nonetheless, and despite vigorous ethical discussion in regular media \citep{zhou_students_2020}, blog posts \citep{torino_data_2020}, and on social media, the ethics of emerging OP technologies has received limited scholarly analysis (cf. \citeauthor{swauger_our_2020} [\citeyear{swauger_our_2020}]). Although moral assessments can be informed by empirical data about online and in-person proctoring --- such as data about test-taker behavior \citep{rios_online_2017} and grade comparisons \citep{goedl_study_2020} --- moral assessments depend crucially on philosophical analysis. In the following ethical analysis, we identify and critically explore the key moral values of academic integrity, fairness, non-maleficence, transparency, privacy, autonomy, liberty, and trust as they apply to OP technologies. 

Some of these concepts are prominent in the new field of AI ethics \citep{jobin_global_2019}, which is burgeoning as AI moves increasingly into many facets of our lives, including in education. In this paper, we suggest that OP platforms are neither a silver bullet for remote invigilation nor, as some would have it, a completely ``evil'' technology \citep{grajek_educause_2020}. This ethical analysis will help to inform concerned individuals while setting out important ethical considerations for educational institutions who are considering OP platforms, including how they might devise appropriate governance frameworks for their use and remain accountable for their decisions. It will also provide a context for various future empirical investigations of OP technologies. 

The essay is structured as follows. The Philosophical Approach section briefly explains the relevance of the central moral values to the OP debate. The Background section provides relevant context concerning exam invigilation and outlines central technological capabilities of popular OP programs. The Discussion section examines important ethical issues raised by the emergence of OP software. Finally, the Conclusion summarizes the ethical lessons for educational institutions and others.

\section{Philosophical Approach}

This essay employs an analytical philosophical approach which includes the application of a range of moral values and principles. Broadly speaking, the moral values and principles we discuss have a place in the philosophy of education \citep{curren_companion_2003}, in the burgeoning field of AI ethics \citep{lin_robot_2017}, and at the intersection of these two fields. An example of this intersection is the ethics of data analytics in student performance \citep{kitto_practical_2019}. Another example of software that raises ethical issues in education is algorithms that make predictions about student performance \cite{sweeney_next-term_2015} that are used to inform grades \cite{hern_ofquals_2020}. Compared to the field of philosophy of education, which can be traced back to Plato’s The Republic, the field of AI Ethics (and, more broadly, digital ethics) is young and still under development. Nonetheless, academics and various organizations that have weighed in on AI ethics tend to agree on the importance of a number of key moral ideas or principles \citep{jobin_global_2019}.

AI ethics principles have occasionally been criticized for their lack of practical specificity and theorical philosophical rigor and for missing wider issues of social, racial, and economic injustice and power imbalance \citep{kind_term_2020}. Additionally, principles such as fairness may be used in confusingly different ways \cite{mulligan_this_2019}. However, these principles or values provide a starting point for scrutinizing AI as a sociotechnical system even if they require further support and contextual specification. Furthermore, our use of such principles in this paper goes some way toward fleshing them out and specifying their application to a novel, concrete socio-technological case, as well as subsequently linking them (briefly) to wider social issues and trajectories.

In a comprehensive global survey of AI ethics guidelines, Jobin et al. identify the ethical principles of transparency, justice and fairness, non-maleficence, responsibility, privacy, beneficence, freedom and autonomy, trust, sustainability, dignity, and solidarity \citep{jobin_global_2019}. In another recent study, \citeauthor{floridi_ai4peopleethical_2018} highlight the ideas of beneficence, non-maleficence, autonomy, justice, explicability, and accountability  \citep{floridi_ai4peopleethical_2018}. Values such as these feature in many other discussions in the AI Ethics literature, and several are especially relevant to online proctoring.

Specifically, the values and principles we shall explore in this paper are fairness, non-maleficence, transparency, privacy, respect for autonomy, and liberty and trust. In addition, we explore the value of academic integrity, which is more specific to education. We shall briefly introduce these values here, and then return to them in more depth in subsequent sections.

Fairness is commonly referred to in discussions of AI ethics, but is also commonly used in a number of different ways in both this field and in moral philosophy and legal analysis. Thus, concerns about fairness may encompass an absence of illegitimate bias, equity considerations in terms of accessibility and opportunity, treating people as ends in themselves not merely as means, and procedural justice. The concept of fairness, which is a species of justice \citep{rawls_theory_2009}, is sometimes connected in AI Ethics to the values of transparency and accountability \cite{jobin_global_2019}. Transparency can refer to the degree to which the determinations or predictions of AI systems are revealed to relevant parties in ways that those parties prefer and can understand, while accountability refers to the degree to which those owning or deploying AI systems have or assume responsibility and/or liability for their outputs. Although transparency is not necessarily or always an ethical value, it is associated with more basic ethical values such as justice and respect for autonomy sufficiently frequently that it is often treated as a key ethical principle in AI Ethics.

Respect for autonomy, a widely prized modern value the prominence of which goes back to the philosopher Immanuel Kant, unsurprisingly features both in AI ethics and in the philosophy of education \cite{siegel_philosophy_2018}. Also holding a prominent place in education is the value of academic integrity, which requires the preservation and nourishment of conditions in which honest and genuine teaching and learning can take place. Non-maleficence is a principle that cautions against doing harm to others. Privacy is relevant to AI ethics because new technologies often collect, process, retain, and interpret vast amounts of often personal and sensitive data. Finally, the values of liberty and trust are increasingly important in ethical discussions of and rising public concerns about the intersection of technology and data gathering and surveillance \citep{zuboff_big_2015}. Values or principles like privacy, fairness, respect for autonomy, transparency, accountability, and trust are also clearly relevant to the treatment of students by universities. Furthermore, some of these values and principles are also implicated in the civic responsibilities and cultural roles of universities. As we shall see, these moral concepts help to illuminate the ethics of OP technologies.

\section{Background}
\label{sec:background}

Digital technologies are used in education in a number of ways. Plagiarism detection tools like Turnitin are widely available and have significantly increased the ability of instructors to uncover academic dishonesty and to teach good academic practice to students. AI teaching systems are emerging that can adapt to the needs and learning styles of individual pupils \cite{bartneck_introduction_2021}. More controversially, AI-based predictions of student performance have been used as inputs into summative grades for students following exam cancellations during COVID-19 lockdowns \citep{simonite_meet_2020}. Similarly, current circumstances have increased the attraction of using digital technologies for remote proctoring of exams. 

Examinations have a long history in both the West and the East. Written public examinations first took place in Imperial China. Centuries later, exams in academia became established in British universities \citep{kellaghan_brief_2019}; their advent in the 1800s gave rise to the first institutional invigilators. Today’s proctors, who possess varying standards of professionalism and expertise \citep{rios_online_2017}, may also be employed by specialist agencies. Proctors also support stressed students \citep{sloboda_combating_1990} and provide equitable exam environments. They are thus required to meet some of the ethical obligations of educational institutions regarding fair and equitable academic assessment. Although not all instructors use exams as an assessment technique, exams still enjoy wide support in teaching \citep{butler_testing_2007} and are likely to persist in the foreseeable future. One reason for the ongoing reliance on exams is that unlike coursework they can be readily invigilated. Instructors can therefore have greater confidence that the work is the student’s own.

Recent technological advances have ushered in software that can be easily integrated into existing university learning management systems and that can arguably assist or even replace live proctors. Reports of the exam session generated by the technology can then be uploaded to a dashboard for convenient review. Although different OP platforms perform broadly similar functions, they sometimes differ, e.g. in the extent of their functionality or in the manner of their use of human invigilators from universities or, alternatively, OP companies. Given this variety and flexibility between and within various OP platforms, we will content ourselves with describing below the more important and/or ubiquitous features of OP technologies. Obviously, their capabilities may increase in time, potentially raising new ethical issues.

\subsection{Monitoring and Control of Devices}

Typically, students must download OP programs or install a web browser extension (which may be deleted post-exam) and permit the commandeering of their computer’s microphone and camera. Different programs allow different degrees of monitoring. They can variously capture screen images, access web page content, block browser tabs, analyze keyboard strokes, and change privacy settings \cite{norman_online_2020}..

\subsection{Candidate Authentication}

OP software can record IP addresses, names, and email addresses, and can request a password or ask other questions to verify candidates’ identity. Programs typically require candidates to display an officially recognized ID card and photo to be matched against their faces by a live proctor (or, conceivably, an AI algorithm). Some programs can analyze the keystroke cadence of typed names to yield biometric substitutes for a handwritten signatures; one program can even request biometrics like fingerprints \citep{examity_auto_2020}. Programs offering more ID data point checks may improve reliability of authentication, while those offering fewer checks may be championed by purveyors as less privacy intrusive.

\subsection{AI-based and Human Online Proctoring}

Online exam invigilation by algorithms and/or a person raise some of the strongest concerns. Examinees may be prompted to activate their webcam and turn their device around 360° to ``scan'' the room for unauthorized materials and family, friends, or housemates \cite{examity_auto_2020,proctorio_comprehensive_2020}. Some programs can detect other devices like mobile phones. The face and body of the candidate can also be monitored, either by means of automated or live human proctoring. 

Some AI algorithms can conduct voice and facial recognition but more commonly perform facial detection and analysis. Machine learning algorithms can be trained on thousands of video examples to recognize movements of eyes and head that appear to correlate with suspicious behavior, like repeatedly glancing away from the screen. The OP system then raises ``red flags'' that an authorized person can review to determine misconduct --- either during the exam (allowing the human invigilator to immediately intervene) or afterwards. Some OP companies claim that the combination of AI and trained human proctors provides greatest accuracy and reliability:

\begin{quote}
The exciting thing about innovating with machine learning technology is that our system is continuously learning, adapting and getting smarter with every exam. ProctorU’s goal in introducing AI into proctoring is not to replace humans but, rather, to strengthen the accuracy of proctoring by assisting humans in identifying details such as shadows, whispers or low sound levels, reflections, etc., that may otherwise go unnoticed \cite{proctoru_harnessing_2020}.
\end{quote}

Purveyors claim that well-designed AI can also mitigate human bias and error \cite{proctorio_comprehensive_2020} and surpass the human ability to accurately detect cheating. Video and audio recordings and analyses are typically stored for a period of weeks or months on company-owned or other servers before being deleted.

\section{Discussion}

As can be seen from this overview, OP technologies have many automated capabilities and, in addition, can readily facilitate remote human invigilation. We should stress that the ethical issues discussed in the present section may pertain to some OP functions but not to others. Furthermore, the technology may give institutions discretion over which capabilities are used. After discussing academic integrity, we examine fairness, non-maleficence, transparency, privacy, autonomy, liberty, and trust as they apply to OP technology. We touch on accountability in the closing section. Violations of one principle and its consequences can sometimes overlap with violations of other principles and their consequences. For example, violations of privacy may cause certain harms and so also be instances of violations of non-maleficence. Table~\ref{tab:principles} summarizes these values and principles and their possible implications for our case.

\begin{table}[!th]
\begin{tabular}{lp{12cm}}
\toprule
\textbf{Value/Principle} & \textbf{Implications for OP exam
technology}\tabularnewline
\midrule
\emph{Academic integrity} & Ensuring academic honesty, rigor,
excellence, and institutional reputation\tabularnewline[2mm]
\emph{Fairness} & Equitable access to technology and remote exam settings\tabularnewline
& Equal, not biased or discriminatory, determination of cheating\tabularnewline[2mm]
\emph{Transparency} &  Transparent use and explanation of the nature of the technology and its
selected functions\tabularnewline
& Transparent use of AI-based ``red flags''\tabularnewline[2mm]
\emph{Non-maleficence} & Effective and safe application of the
technology which does not cause harm to the subject\tabularnewline[2mm]
\emph{Privacy} & Privacy in collection and security of personal data and
exposure of body, behavior, and home spaces\tabularnewline[2mm]
\emph{Respect for autonomy} & Examinee autonomy regarding personal data
use, use of AI, video recordings, strangers as proctors\tabularnewline[2mm]
\emph{Liberty and trust} & Potential wider effects on freedoms, use of
digital technologies, and society's trust in AI, universities,
etc.\tabularnewline[2mm]
\emph{Accountability} & Accountability by the entity using the
technology for misuse and processes for individuals to contest wrongful
outcomes\tabularnewline
\bottomrule
\end{tabular}
\caption{Ethical values/principles and their implications for
OP exam technology}
\label{tab:principles}
\end{table}

\subsection{Academic Integrity}

Academic integrity, a vital quality and value in academia, can be threatened by student ignorance, dishonesty, and misconduct. Forms of academic dishonesty and misconduct include impersonation, unauthorized use of cheat notes, and the copying of exam answers from fellow students or online sites. OP programs target all these illicit activities. There are several ethical reasons why it is vital to prevent academic dishonesty \citep{kaufman_moral_2008}, which we can briefly enumerate. First, the value and viability of courses and universities depend on their academic integrity and educational rigor. Second, permitting cheating is unfair on students who are academically honest. Third, knowledge that others are cheating can create for honest students an invidious moral choice between self-interest (e.g. where class rankings matter) and personal integrity, and a hurtful sense of both being taken advantage of by fellow students and let down by the university. Fourth, universities arguably bind themselves to providing students with (in some sense) a moral education alongside an intellectual education, minimally by nourishing a favorable academic culture in which academic integrity is salient \citep{dyer_academic_2020}. 

Although universities rightly encourage in students an autonomous and sincere commitment to honest behavior, failure to invigilate where necessary to prevent cheating above a certain level can, in addition to other consequences, convey the impression that academic honesty is unimportant. What that level is in any particular case requires a difficult judgement. Yet although its effects may be hard to judge, it would be too quick to simply dismiss the idea that student cheating can sometimes have corrosive effects on academic integrity and all that entails. Some studies suggest that students more often cheat in online testing environments than traditional exam rooms \citep{srikanth_modern_2014}, although there are conflicting views \cite{stuber-mcewen_point_2009}. Cheating may help students to achieve higher grades in assessments, but it may also degrade their learning and longer-term interests. For all these reasons, both universities and students have significant interests in the maintenance of academic integrity. Consequently, the need to preserve academic integrity constitutes a non-trivial reason for considering the use of OP technologies.

\subsection{Fairness --- Equity and Accessibility in AI use}

Remote invigilation via OP technologies promises accessibility benefits for students who normally study on campuses. Institutions may save on costs of hiring exam centers and professional invigilators, which is important in the face of severe budgetary constraints exacerbated by COVID-19 lockdowns\footnote{Note however that OP technology can cost many thousands of dollars for institutions \citep{grajek_educause_2020}.}.  OP can facilitate a greater range of online course offerings and benefit students from distant places or with limited mobility. The technology allows exams to be scheduled day or night, which could particularly benefit parents. For such reasons, OP may help promote more equitable outcomes, including for traditionally excluded social groups. 

However, some students may be disadvantaged by OP in certain ways. This includes students who lack reliable internet connections --- an online exam may be voided if the internet even momentarily disconnects. Some students lack appropriate devices, such as web cameras, and home or other environments in which to sit exams. To be fair, OP providers largely appear eager to ensure that their programs are executable on numerous devices and do not require super-fast internet connections. Also, universities may loan devices and arrange for select students to sit exams on-campus or at other locations. A potential drawback in some such cases is that doing an exam later than the rest of the cohort may delay course progression or graduation. Hence, there are logistical issues, with fairness implications, for institutions to consider.

\subsection{Fairness --- Bias, Discrimination, and AI-facilitated Determination of Cheating}

OP platforms using AI and/or live proctors may increase fairness for honest students by identifying relatively more cases of cheating. Some proponents claim that digital proctoring does better on this score than traditional invigilation where the ratio of proctors to test-takers is very low \cite{dimeo_online_2017}. This claim, of course, would need to be backed by empirical studies. In addition, the potential for OP’s to create unfairness also needs to be considered. Unfairness can relate to inequitable outcomes and unjust processes. This may play out in different ways. For example, false negative identification of cheating may constitute unfairness for non-cheating students, while false positives may result in unfairness for those examinees. Unfairness may flow from the use of AI, from remote human invigilators, or from both together. This needs to be spelt out.

The general fairness problems created by the use of AI and ML are the subject of vigorous contemporary public and academic discussion in AI ethics. Deployment of ML has starkly exposed its potential for inaccuracy. For example, facial recognition technology has been criticized as inaccurate, and has even resulted in legal action, despite the fact that the ML algorithms may have been trained on thousands or millions of images \cite{peters_ibm_2020}. Energetic debate has similarly centered on the so-called biases that can afflict ML. Again, facial recognition software has been associated with bias in the (mis)recognition of certain racial groups and gender \cite{buolamwini_gender_2018}. Notorious cases of inaccuracy and bias in ML include automated reviews of curriculum vitae for job applications which favor male candidates, determinations of parole conditions for offenders which apparently discriminate against people of color, and the disproportionate allocation of policing to disadvantaged communities \cite{oneil_weapons_2016}. Thus, machine bias can create unfairness \cite{jobin_global_2019}. Another form of bias or discrimination may arise through the model used by the OP for ``normal'' or ``acceptable'' exam behavior. OP providers refer to flagging suspicious gestures and even tracking eye movements. Yet people with disabilities or neuro-atypical features may not always behave in a way that is recognized by these processes \citep{swauger_our_2020}, and this may lead to false positives as such people are red flagged for cheating through the manifestation of their normal behaviors.  

Bias can creep into ML through input of skewed and poorly representative training data or through the mechanisms of pattern-searching \cite{mehrabi_survey_2019}. Presumably, this could occur in the training and operation of ML in online cheating analysis. As OP platforms accumulate increasingly larger data sets on which to train, their reliability should increase. But bias and inaccuracy may never be fully eliminated, and some forms of unfairness may not be solvable by purely technical means \citep{selbst_fairness_2019}, leaving the potential for students to be unfairly charged with cheating. Nevertheless, unfairness in socio-technical systems need not always be the outcome of ML bias. OP companies stress that it is, after all, not the AI algorithm that ultimately makes a judgement about academic dishonesty, but a knowledgeable human being, such as the course instructor. Furthermore, instructors may choose which settings they will and won’t use --- for instance, they might choose to disable or ignore AI algorithms that track eye movements. 

While true, this flexibility does not totally eliminate ethical concerns. For example, instructors may have unwarranted faith in the red flags, such as the automated flagging of ``suspicious'' head movements. The problem is magnified when we consider the conscious and unconscious inclination for some people to over-trust AI \citep{dreyfus_mind_2000}. Even where psychological bias is absent, instructors may be unsure how to interpret some red flags and may draw incorrect inferences from them. Certain flagged events, such as when the test-taker is replaced with another person, are relatively easy to assess. But more subtle flags may be much harder to appraise, such as flags for ``low audible voices, slight lighting variations, and other behavioral cues'' \citep{proctoru_proctoru_2020}. Further, if the ML element is intended to enhance detection of cheating over and above a human observer carefully attending to the same images (etc.), then it follows that an independent level of trust is intended to be invested in the AI assessment. As ML technology advances, greater epistemic weight will likely be placed on its judgements.

\subsection{Non-maleficence}

Reliance on OP technology raises risks of harm for both students and universities. As mentioned, there is a risk to students of false claims of cheating that did not occur. Wrongful allegations of academic misconduct, especially where there is no process for contestability, may affect job prospects, self-confidence, and personal trust in the university. For universities, false negatives and positives could more broadly undermine social trust in the integrity of the institution\footnote{We discuss trust further below.}.  Therefore, it is important that such systems are effective, that they work with a sufficient degree of accuracy, and that there is clarity about their reliability. But, as we have noted, the operation of such systems is often opaque, and although claims are made about accuracy, the OP websites rarely or never cite rigorous studies to justify their claims and to eliminate concerns about false positives; e.g.\ \cite{examity_auto_2020,proctorio_comprehensive_2020}. 

One could imagine hypothetical situations that clearly involve unfairness and harm related to assumed belief, group membership, or behaviour. Suppose, for example, that some future AI proctoring system red flags the presence in the examinee’s room of white supremacist propaganda or pornographic material; or suppose the AI system is biased towards red-flagging suspicious eye movements in people with disabilities, or assumes that black students need closer monitoring than white students. Again, imagine the program casts doubt on students’ honesty purely from a brief unintelligible exchange of words with someone who happens to enter the room. 

Whether AI-led or human-led, post-exam determinations of cheating differ from in-person or live remote invigilation, where the primary anti-cheating mechanism is typically to warn students at the precise time of the potential infraction (e.g. when students are seen conversing). Unlike subsequent review of captured OP data, that mechanism does not depend on an official charge of academic dishonesty, but on its immediate prevention. Some test-takers may simply have idiosyncratic exam-taking styles, or disabilities and impairments, that trigger specious AI red flags. Even falsely suggesting that these individuals are academically dishonest, let alone accusing and penalizing them, would potentially be unfair and harmful. Even though such a false suggestion or imputation is less morally serious than a false official condemnation, they are still morally serious. Further, recipients of spurious insinuations are likely to receive them as an injustice and to feel corresponding hurt. In addition, such individuals, in an effort to avoid this potentially wrongful treatment, may be forced to disclose personal idiosyncrasies or impairments, compromising their privacy and potentially doing them harm in the process.

\subsection{Transparency}

Uncertainty may persist about how precisely the AI identifies ``cheating behavior.'' Some OP websites are more transparent than others about how their AI systems work. But even with some explanation it can be confusing and difficult to gain an adequate understanding of how they compute red flags and how reliable those determinations are\footnote{For one example of an attempt to explain how AI-based judgements are made using a ``credibility index''; see \citeauthor{mettl_mettl_2018} [\citeyear{mettl_mettl_2018}]. This program allows users to select which indicators or patterns (e.g. eye movements) to incorporate and which to exclude from the AI analysis, as well as the weight they carry.}.  One representative explains that their company uses an:

\begin{quote}
incredibly futuristic AI Algorithm that auto-flags a variety of suspicious cases with 95\%+ accuracy \ldots With AI-driven proctoring, the algorithms will soon become trained enough to prevent cheating 100\%, a guarantee that a physical invigilator cannot always promise \cite{kanchan_top_2019}.
\end{quote}

Compared to, perhaps, the plagiarism detection tool Turnitin, proctoring AI may strike users as highly opaque \citep{castelvecchi_can_2016}. The problem of AI ``black boxes'' is one reason why ethicists stress the moral need for transparency in AI \citep{reddy_governance_2020}. Transparency in this context may work at different levels and different times. At the outset, students need to understand enough about the OP process to know what is expected of them in exams so as not to trigger a red flag. Students will also need information on how to contest any adverse finding and their rights of appeal, and those who are accused of cheating will need to know the basis on which that allegation is made. 

Admittedly, not all of this information need be presented to students upfront, especially given concerns about information overload and about students gaming the system. Nonetheless, academic fairness requires that the evidence and procedures on which accusations of cheating are made are defensible and transparent. To reduce the risks of unfairness and emotional harm, OP companies and universities should be transparent about how the technology works, how it will be used in particular circumstances, and how it will impact on students, including those with idiosyncrasies and disabilities.

\subsection{Privacy}

OP technologies raise moral concerns about privacy which privacy laws, and university policies and governance, may not adequately address \citep{pardo_ethical_2014} --- especially given that many jurisdictions have privacy laws that have not been amended to adjust to the data collecting capacities of new digital technologies \citep{australian_government_oaic_2020}. OP technologies collect a number of kinds of data, including the capturing and storage of information from devices, the gathering of biometric and other ID details, and the audio and video recording of students and their environment. It should not be surprising that some students have a sense of ``Big Brother invading their computers'' \citep{dimeo_online_2017}.

Privacy is a large philosophical topic. A rough distinction can be made between private and public domains \citep{warren_right_1890}. Privacy can relate to the (non)exposure to other individuals of one’s personal information and one’s body, activities, belongings, and conversation \citep{gavison_privacy_1980,moore_privacy_2003}. However, what is private for one person may, in a recognizable sense, not be private for another \citep{moore_privacy_2015}. For example, I may strongly prefer that no strangers gaze inside my bedroom and watch me studying; whereas you, who do not draw your curtains, may not care. Exposure of my bedroom and activities to passers-by represents in my case a loss of privacy; in your case it does not. So, there is an intelligible sense in which the determination of privacy and its breaching can turn partly on individual perspectives about the personal. At the very least, we can say that the moral seriousness of exposure is plausibly related, to some extent, to these individual preferences. 

While the moral ``right'' to privacy may sometimes justifiably be infringed (e.g. in law enforcement) it is still a vital right. For some philosophers, privacy’s value essentially reduces to the value of liberty and autonomy \citep{thompson_nature_1979}, i.e. to a person’s ability to act and make choices about their own lives. For others, its importance relates to possible harms resulting from public exposure of the personal \citep{rachels_why_1975}, such as social embarrassment and deleterious financial or employment repercussions. We might regard privacy’s importance not as confined to a single philosophical conception, but to a range of conceptions that cover respect for autonomy, the causation of various kinds of harm, and so on.

OP technologies may threaten personal privacy in several ways. Reports exist of inadvertent capture of personal or sensitive information, such as in one case a student’s credit card details that were accidentally displayed on their computer screen \citep{chin_exam_2020}. While technology designers might address some such risks, there are additional risks concerning data security. Captured information can be stored in encrypted form on host servers such as Microsoft and Amazon servers. For their part, many OP companies claim that they have no access to this encrypted information and therefore cannot view video recordings or obtain sensitive personal data. Furthermore, purveyors claim to be compliant with legal protections like the EU’s General Data Protection Regulation (GDPR), which carries heavy penalties for breach. 

Companies may also have internal rules against sharing data with third parties and for commercial gain \citep{dennien_uq_2020}, and universities too are required to have stringent cybersecurity and privacy policies. However, there can be no absolute guarantee against leakage of data or successful cyberattacks on servers used by companies or universities. The maintenance of such privacy is never completely certain: these kinds of cyber risks are always present with any data collected by any institution \citep{anu_anu_2019}. It is nonetheless possible that students may feel particularly anxious about the possible loss of the kinds of sensitive personal information (e.g. video recordings, certain data from personal computers) collected by OP technologies.

As we saw, some people worry that OP platforms are especially intrusive because they readily facilitate video (and audio) capture of examinees and its live or subsequent review by a person. This concern may be countered by proponents of OP technologies as follows. Students necessarily relinquish aspects of their privacy in education. In-person invigilation, which is morally uncontroversial, is already privacy-invasive: strangers or instructors watch students like hawks, scrutinizing their activities and personal belongings. On this moral view, online proctoring is essentially the same in an ethical sense as in-person invigilation. We may start by noting that students who have used Examity say that

\begin{quote}
    ``it feels much weirder than proctoring with a professor \ldots They’re being watched closer up, by a stranger, and in a place more private than a classroom…students described their experiences as everything from `uncomfortable' to `intrusive' to `sketchy''' \citep{chin_exam_2020}.
\end{quote}

One element of the privacy intrusion relates to human invigilators seeing into the home environment of the student, such as bedrooms or lounge rooms. Another element is that of other human beings watching video of the faces and upper bodies of students themselves. To make the analogy with traditional invigilation more truly comparable, then, we must imagine an in-person supervisor sitting near the examinee and staring at them throughout the exam. Such observation would include scrutinizing the student’s expressions or micro-expressions, perhaps with the help of an AI device. 

Furthermore, OP may allow the human invigilator, who may reside locally or on the other side of the world, to re-watch the video and use its pause function, potentially in private. In traditional exam rooms, the presence of other students and instructors provides a degree of security and protection. In contrast, the student who is invigilated online cannot know, even when they are given assurances by universities and OP companies, how the online human proctor uses the video. For example, students cannot be sure that they are not being leered at that online proctors have not shared their images shared with third parties. The online scenario should strike us as potentially more invasive of privacy than in-person invigilation, irrespective of whether students --- including students whose histories and psychologies render them particularly averse to being closely watched by strangers --- have the additional concern that viewers may take a prurient interest in them. Besides, some students evidently have that view. Further, because (as we suggested) what constitutes a loss of privacy turns in one sense partly on the individual’s own perspective, OP in those cases is more invasive of privacy. It follows that the hurdle for justifying its use is that much higher.      

\subsection{Respect for Autonomy}

Autonomy might be restricted by online proctoring in a number of ways. For example, it may require students to avoid doing things they can often do in traditional exams, such as muttering to themselves, looking to the side, going to the bathroom, etc. --- lest they raise automated red flags about suspicious behavior. Some students may simply prefer not to be invigilated by AI or by online human proctors, or to have their images and personal data collected and viewed. Philosophers often regard respect for autonomy as a fundamental ethical value \citep{christman_autonomy_2018}. Autonomy in this sense implies self-governance, or the ability of a rational and mature agent to form and act on decisions and personal values free of compulsion and manipulation. As we saw, some philosophers ground the value of privacy in the value of autonomy. 

We should be clear, however, that respect for autonomy is not reducible to respecting privacy: respect for autonomy can apply to the use of personal information even where loss of privacy is not at stake (e.g. because data is anonymized). Further, respecting autonomy may require providing agents with opportunities for informed consent. For informed consent to apply, the choice must be made voluntarily and exercised with liberty and without coercion, and the chooser must have adequate knowledge of the nature, risks, and benefits of committing to or refraining from the relevant action (cf. \citeauthor{clifford_consumer_2020} [\citeyear{clifford_consumer_2020}]). This requires transparency about the nature and potential effects of OP programs. A genuinely robust standard of consent would also allow students to be able not to consent to OP and to choose instead a human invigilator. From the perspective of a university, this kind of discretion granted to students may be unmanageable; but the point emphasizes the need for other processes to protect students’ interests. 

It might be observed that autonomy, and the prima facie requirement for informed consent, are already justifiably restricted in education. Educational limitations on liberty extend, quite obviously, to the prevention of cheating and much more (as when personal student information is collected for enrolment). One early student criticism of the plagiarism tool Turnitin likened its use to the coercive drug testing of students \citep{glod_students_2006}, but such moral objections are now often (though not universally) regarded as exaggerated. Indeed, our attitudes towards novel technologies can change with increased familiarity and understanding. However, deciding when it is justified to limit autonomy for the sake of academic integrity requires moral (and not just legal) judgement.

Most ethicists, including that classic defender of liberty John Stuart Mill, acknowledge that coercion and compulsion are sometimes justified, most obviously when the exercise of a freedom is likely to result in significant harm to others \citep{gaus_liberalism_2020,mill_liberty_1966}. But even though and when that is so, respect for autonomy may imply that limitations upon autonomy be minimized wherever possible, and that relevant information be provided transparently to students who are being compelled by their universities. This would include information related to the above concerns, along with other concerns. For example, purveyors of OP technologies may use data derived from student exams to train ML algorithms \citep{proctoru_proctoru_2020} without the students being (adequately) informed that their data will be so used. Such use may arguably produce good outcomes (e.g. improving accuracy and reducing bias in AI proctoring), but institutions which fail to investigate data use arrangements and/or inform students accordingly could be said to thereby disrespect the autonomy of those students.  

\subsection{Liberty and Trust}

To conclude our investigation of the ethical issues raised by OP technologies, we shall briefly discuss some possible wider ethical risks. Although these risks are admittedly much less certain than those explored above, they are real enough to warrant taking into account when forming a comprehensive ethical judgement about this emerging socio-technical example. The above concerns about the intrusive and invasive nature of OP technologies have a possible connection to broader technological and social changes and trajectories. These trajectories may include increased surveillance and the step-by-step evolution of a security state \citep{reiman_driving_1995}, new risks of personal data being publicly ``reidentified'' despite claims it is anonymised \citep{culnane_health_2017}, a constriction of the private domain \citep{nissenbaum_privacy_2009}, and the spread of AI-based decision making in ways that some consider problematic. Consider, for example, contemporary public perceptions arising from the use of AI facial recognition, the scraping and dubious employment of personal data from social media, tracking and tracing during the COVID-19 pandemic, AI decision-making in jurisprudence, and so on \citep{feldstein_global_2019}\footnote{For example, a website for ``Proctortrack'' says that its ``RemoteDesk'' solution goes beyond exam proctoring to provide automated monitoring of people who work from home'' \citep{kanchan_top_2019}.}.  

It is at least arguable that OP technology could (even if modestly) contribute to these worrying social trajectories or possibilities. Such risks are, as we stress, very difficult to assess, but that does not mean they may be ignored. Of course, some universities have chosen not to use OP technology. Furthermore, we can perhaps expect that many universities will make diligent efforts to protect and foster respect for privacy, liberty, autonomy, and fairness \citep{kristjansson_emotions_2017} when they use these technologies. Such efforts would, of course, be entirely proper. Indeed, it may be suggested (though we cannot argue it here) that universities should recognize and reaffirm their standing as bulwarks against the natural proclivity of governments and powerful corporations to intrude into people’s private lives and to chip away, deliberately or unthinkingly, at their freedoms. Some students and university staff evidently feel that OP platforms could damage a university’s ``culture of trust.'' Such an effect could have wider reverberations. That is a reason for taking the ethical aspects of OP technologies seriously.

The weight of the above concerns will depend not only on cultural factors and differences but also partly on factors such as the extent of opposition to OP technology amongst students and staff, and the relative intrusiveness of the various proctoring functions which are developed by companies and chosen by institutions. We have already discussed a range of more or less ``invasive'' and ``intrusive'' capabilities that, for some people, have vaguely Big Brother overtones. Disquiet would mount still further if OP platforms allowed, say, facial recognition, the undisclosed on-selling of test-taker data, and use of AI to generate Uber-style ratings to indicate an examinee’s honesty while closing off avenues for appeal and rebuttal. 

In today’s digital and cultural climate, none of these further possibilities may be dismissed out of hand. At some point along this line, universities may, in light of their social responsibilities, want to take a stand against not only the ethical risks to students and to their own reputations, but against the risks created for society more generally of endorsing particularly invasive or intrusive technologies. In any case, the point we are underlining here is that OP technologies need to be considered not just from the perspective of their potential immediate and local effects, but also from the perspective of their more distant, wider, and longer-term potential effects, even if those effects are much harder to measure and predict with any certainty.

\section{Conclusion: Justification and Accountability}
\label{sec:conc}

Debate and disagreement about the appropriateness of remote OP technologies in distance education, and in circumstances like the ongoing COVID-19 pandemic, is bound to continue. As we saw, there are considerations that speak in favor of OP technologies despite their drawbacks. Indeed, it is fair to acknowledge here that in-person proctoring is not ethically perfect either: it can both miss cheating and similarly result in unfair accusations of academic dishonesty. Furthermore, we have accepted that it is vital to maintain academic integrity to protect both students and institutions. It is true that the pedagogical value of high-stake examinations is sometimes questioned\footnote{Grajeck reports: ``Three in ten institutions are considering broad changes to assessment. Exams are common, but they are only one way to assess learning. The [COVID-19] pandemic is providing 31\% of institutions the opportunity to consider more authentic demonstrations of knowledge and skills'' \citep{grajek_educause_2020}.};  and faced with the ethical problems of OP technology some academics will adopt alternative assessments. But, on the assumption that high-stake exams have value and will persist in education, there are reasons for regarding at least some OP technologies and capabilities as representing acceptable ``proctors'' or proctors’ assistants.

Nonetheless, the above analysis revealed that OP platforms raise ethical concerns over-and-above those affecting live and in-person exam invigilation. These concerns include an uncertain risk of academic unfairness associated with AI-informed judgement, further diminution of student privacy and autonomy, and (perhaps) increased distrust towards institutions that are bastions of social values. Another fear, partially dependent on these former fears, is that OP platforms could contribute to the social trajectories of growing surveillance, liberty and privacy loss, mining of massed personal data, and dubious instances of AI decision-making. 

It is difficult to know whether the benefits of OP technologies outweigh their risks. The most reasonable conclusion we can reach at the present time is that the ethical justification of OP technologies and their various capabilities requires balancing as best we can the concerns with the possible benefits. In any case, it behooves educational institutions to consider the ethical considerations we have explored in this paper, and, should they choose to adopt OP technologies, to accommodate those considerations in their policies and governance plans. Those institutions need to have the right systems in place to remain accountable for such choices. 

Accountability as a value and principle would require that students are adequately and transparently informed about the impacts of the capabilities of any particular OP technology that is selected on the ways in which cheating is determined and privacy potentially affected. Accountability would also require systems for addressing potential injustices. Future empirical investigations may further illuminate the effects of OP technologies and their widespread uptake on issues like fairness, privacy, harm, autonomy, trust, liberty, and accountability.

\paragraph{Funding} The authors received no funding for this paper.

\paragraph{Conflicts of interest/Competing interests} Nil

\paragraph{Availability of data and material} N/A

\paragraph{Code availability} N/A

\paragraph{Declaration of interest} No conflict of interest exists in the submission of the manuscript. The work described is original research that has not been published previously and not under consideration for publication elsewhere, in whole or in part.

\bibliography{references}
\bibliographystyle{plainnat}

\end{document}